%% file: main.tex
\documentclass[lettersize,journal]{IEEEtran}
\usepackage{amsmath,amsfonts}
\usepackage{algorithm}
\usepackage{array}
\usepackage[caption=false,font=normalsize,labelfont=sf,textfont=sf]{subfig}
\usepackage{textcomp}
\usepackage{stfloats}
\usepackage{url}
\usepackage{verbatim}
\usepackage{graphicx}
\usepackage{cite}
\usepackage{enumitem}
\usepackage{algorithm}
\usepackage{algpseudocode}
\usepackage[normalem]{ulem}
\usepackage{multirow}
\usepackage{colortbl}
\usepackage{hhline}
\usepackage{float}

\newcommand{\ie}{\emph{i.e., }}
\newcommand{\eg}{\emph{e.g., }}

\newcommand{\etc}{\emph{etc.}}
\newcommand{\wrt}{\emph{w.r.t. }}

\newcommand{\aka}{\emph{aka. }}

\setlist[itemize]{leftmargin=4mm}

\begin{document}

\title{ToDA: Target-oriented Diffusion Attacker against Recommendation System}

\author{Xiaohao~Liu,
        Zhulin~Tao*,
        Ting~Jiang,
        He~Chang,
        Yunshan~Ma,
        Yinwei~Wei,
        Xiang~Wang

\thanks{
X. Liu, Z. Tao, T. Jiang, and H. Chang are with the State Key Laboratory of Media Convergence and Communication, the Communication University of China (e-mail: taozhulin@gmail.com).
Y. Ma is with the National University of Singapore (e-mail: yunshan.ma@u.nus.edu).
Y. Wei is with the Monash University (e-mail: weiyinwei@hotmail.com).
X. Wang is with the University of Science and Technology of China and Institute of Artificial Intelligence, Institute of Dataspace, Hefei Comprehensive National Science Center (email: xiangwang1223@gmail.com).
\\
}
}

\markboth{Journal of \LaTeX\ Class Files,~Vol.~14, No.~8, August~2021}%
{Shell \MakeLowercase{\textit{et al.}}: A Sample Article Using IEEEtran.cls for IEEE Journals}


\maketitle

\begin{abstract}
Recommendation system (RS) has become an indispensable tool to address information overload, simultaneously enhancing user experiences and bolstering platforms' revenues. However, due to the public accessibility, it is susceptible to specific malicious attacks where attackers can manipulate user profiles, leading to biased recommendations, \aka \emph{shilling attacks}.
Recent research uses generative models and integrates additional modules to craft deceptive user profiles, ensuring they are imperceptible while causing the intended harm. 
Despite the effectiveness, these models face challenges of learning dilemmas and inflexibility, which can lead to suboptimal performance.

In this paper, we propose a novel \textbf{T}arget-\textbf{o}riented \textbf{D}iffusion \textbf{A}ttack model (ToDA), pioneering the investigation of the potential of diffusion models (DMs).
DMs have showcased remarkable capabilities in areas like image synthesis, recommendation systems, and adversarial attacks, providing finer control over the generation process.
To assimilate DMs within shilling attacks, we address their inherent benign nature and the narrowness of the local view.
ToDA incorporates a pre-trained autoencoder that transforms user profiles into a high-dimensional space, paired with a Latent Diffusion Attacker (LDA). 
LDA introduces noise into the profiles within the latent space, adeptly steering the approximation towards targeted items through cross-attention mechanisms.
The global view, implemented by a bipartite graph, enables LDA to extend the generation beyond the on-processing user feature itself and bridges the gap between diffused user features and target item features.
Extensive experiments compared to several SOTA baselines  demonstrate ToDA's efficiency and efficacy, highlighting its potential in both DMs and shilling attacks.

\end{abstract}


\input{sections/1-intro}

\input{sections/3-Formulation}

\input{sections/4-Method}

\input{sections/5-Experiment}

\input{sections/2-Related.tex.bak}
\input{sections/6-Conclusion}

\bibliographystyle{ieeetr}
\bibliography{sample-base}

\input{sections/Appendix}




\end{document}

%% file: sections/1-intro.tex
\section{Introduction}

Recommendation system (RS) primarily endeavors to capture users' preferences through their historical interactions, thereby predicting potential item candidates that would likely attract users~\cite{BPR, SVD++, LightGCN}. As an effective countermeasure against information overload, it remarkably enhances the user experience while concurrently boosting the revenue of merchants in many web services (\eg e-commerce~\cite{CrossCBR} and content-sharing platform~\cite{MMGCN}). 

\begin{figure}
    \centering
    \includegraphics[width=0.99\linewidth]{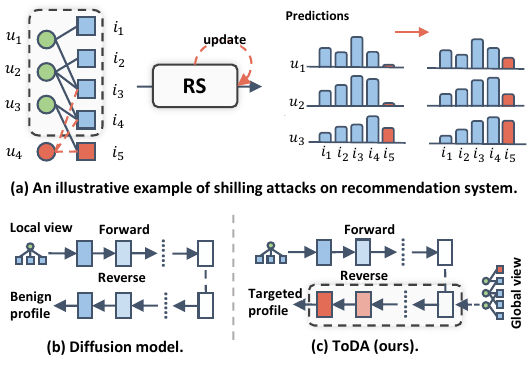}
    \vspace{-4mm}
    \caption{Illustration of a shilling attack example and the comparison between conventional diffusion models and ToDA, which derives the targeted profile by leveraging global view graph and target information.}
    \label{fig:shilling_attack}
    \vspace{-4mm}

\end{figure}
    
In light of the ubiquitous nature and public accessibility of recommendation systems (\ie recommendation models are trained based on user data which is usually accessible), the necessity for prioritizing security has become increasingly apparent~\cite{TrustworthyRS, AdversarialRS}. 
Although many works leverage adversarial learning with disturbed samples to improve the robustness~\cite{FNCF, ACAE, AMR}, recommendation system is still vulnerable to malicious attacks~\cite{PCAttack, AUSH, HRGAN, LegUP,GSPAttack,RAPU-R,WB}. 
In other words, attackers are able to glean interaction histories of users and subsequently construct fabricated user profiles (\ie a sequence of user-item interactions) as inputs to the recommendation system, thereby promoting or demoting the target items. 
This action is recognized as \emph{shilling attack}.
Figure~\ref{fig:shilling_attack} (a) depicts the example of a shilling attack. 
The attacker crafts a fake user $u_4$ who has interacted with three items, $i_3$, $i_4$, and $i_5$ where $i_5$ is the target item. 
With the training of historical and injected interactions, recommendation system updates its parameters, resulting in a prediction shift. 
Specifically, the target items gain a higher probability of being candidates in the recommendation list. 
Obviously, shilling attacks undermine the integrity of recommendation systems, leading to unfair exposure of items, and consequently eroding users' trust. 

Several efforts are devoted to the design of shilling attacks as the security concerns and insights provided to defense-side research. 
The field of shilling attack research has seen a clear evolution.
Initially, 1) researchers primarily adopt heuristic rules to manually craft user profiles~\cite{shillingAttack, bandwagon}
or approximate an optimization problem~\cite{DPAF, PAGR, DLAttack}. 
Afterward, 2) generative models gradually take the dominant role of shilling attacks to the present: 
these methods can be categorized into autoencoder (AE)-based~\cite{PCAttack,RAPU-R}, generative adversarial network (GAN)-based~\cite{TriAttack, GSA-GAN, AUSH, HRGAN, LegUP, GSPAttack}, and reinforcement learning (RL)-based~\cite{ARG, TDP-CP, KGAttack, CopyAttack, LOKI, Poisonrec} strategies, which are in an autoagressive manner. 
To adapt these methods for shilling attack tasks, an additional module is often incorporated to optimize the attack objectives~\cite{RAPU-R, LegUP, TriAttack}. 
Notably, recent shilling attack methods have been entrenched in GANs or autoregressive RL methods for several years since 2020~\cite{AUSH, HRGAN}.
Despite their efficacy, these models face challenges, such as unstable training and the exploration-exploitation dilemma, which lead to suboptimal results.
To break down the standstill of recent shilling attacks and mitigating the inflexibility and learning dilemmas, a new generative paradigm for future shilling attacks is necessary. 

Diffusion models (DMs) have emerged as a cutting-edge technique for generating data across various domains, like computer vision~\cite{DMs_vision_survey} and natural language processing~\cite{Diffuser, Diffuseq}. At their core, these models simulate the process of diffusion, denoising the data step by step to reconstruct the data.
Some works in recommendation system (RS) harness DMs to predict user preference under the noisy scenario~\cite{DiffRec} and forecast the users' tastes evolving over time~\cite{DiffuRec}. 
In adversarial attack field, DMs help to purify samples' perturbations~\cite{DM_purification}, craft malicious visual samples~\cite{DiffAttack} and improve model robustness~\cite{DM-Improves-AT, DensePure}. 
DMs provide finer control over the generation process, allowing for enhanced interpretability and precision. Their inherent denoising capability also ensures the generation of high-quality data.
Given the success of DMs in these areas,  there is growing interest in their potential for shilling attacks, an intriguing area where DMs remain largely unexplored.

Utilizing DMs is natural yet challenging, presenting unique challenges. 
As illustrated in Figure~\ref{fig:shilling_attack} (b), we summarize the following two main challenges.
\begin{itemize}
    \item {The inherent nature of DMs is benign.} 
    They are designed to understand and replicate patterns without any malicious intent. 
    Even though some attempts in RS and adversarial attack showcase the potential, tailoring DMs specifically for shilling attacks requires significant modifications to ensure the generated profiles are both imperceptible and effective in manipulating the victim RS.
    \item {DMs typically have a narrow focus}, often concentrating on a single sample during generation (\ie local view).
    However, shilling attacks necessitate a broader perspective that outwards the on-processing pattern to recognize auxiliary items (\ie global view). 
    For example, attackers must comprehend the interrelationships among various items and their mutual influences to identify a range of items relevant to the target, instead of merely assigning target items to the fabricated profiles.
\end{itemize}
To address the above challenges, we propose a novel \textbf{T}arget-\textbf{o}riented \textbf{D}iffusion \textbf{A}ttack model, termed \textbf{ToDA}.
As shown in Figure~\ref{fig:shilling_attack} (c), ToDA derives targeted profiles step by step during the reverse procedure, where the global view graph and target information are incorporated.

Specifically, ToDA hinges on a latent diffusion model by first encoding user profiles into a high-dimensional space to diminish the computational cost and facilitate model flexibility. 
Therefore, we are able to subtly add noise to latent features at each forward step, and employ an approximator to reconstruct every state, referencing the prior step during the reversion process. 
This procedure allows us to sample natural profiles that conform to victim RS, inherently ensuring the desired imperceptibility.
To steer the latent diffusion model from being malicious, we harness the Target-oriented Approximator within the reversion by using cross-attention to condition the target item's features under the global view. 
To this end, we adopt a bipartite graph (\ie user-item graph), thus making it possible to extend the generation of a broader horizon outwards the on-processing user feature itself, and generating more diverse and relevant target item features.
Wherein, a SOTA GNN encoder in collaborative filtering (\ie LightGNN~\cite{LightGCN}) is adopted to catch the high-order correlations of target items.
Without an extra module or attack objective, we endow the diffusion model with attack ability. 
Our ToDA is simple yet flexible and powerful for shilling attacks. 
We compare ToDA with several classical methods and SOTA generative models in the context of shilling attacks, like LegUP~\cite{LegUP} and GSPAttack~\cite{GSPAttack}, and exploit significant improvements. 
Overall, our contributions are threefold:
\begin{itemize}
    \item We investigate the previous works from a generative standpoint, highlighting the untapped potential of utilizing DMs for shilling attacks. 
    To the best of our knowledge, this is a pioneering effort in assimilating DMs within shilling attacks.
    \item We devise a novel target-oriented diffusion attacker, ToDA, underpinned by the latent diffusion model paired with a target-oriented approximator. 
    We innovatively confer attack ability to DMs, filling a blank in the confluence of DMs and shilling attacks. 
    \item Through extensive experiments, we present a meticulous analysis of ToDA, which empirically demonstrates both its reasonability and effectiveness. 
\end{itemize}

%% file: sections/3-Formulation.tex
\section{Preliminary}

In this section, we elucidate the goals of shilling attacks and determine the attacker capabilities in manipulating these systems, including incomplete data, black-box setting, and no extra knowledge. 
These settings ensure that our proposed attack are adapted to real-world scenarios.
Furthermore, we provide a formal overview of recommendation system.

\subsection{Shilling Attacks}
\subsubsection{Attack goal}

In the context of recommendation systems, shilling attacks represent a deliberate effort to manipulate the recommendation process by injecting fraudulent user profiles into the system. These attacks aim to distort the recommendation outcomes in favor of certain items or to undermine the integrity of the recommendation algorithm. Formally, let $\mathbf{Y} \in \{0,1\}^{n\times m}$ denote the matrix of observed user-item interactions, where $n$ represents the number of users and $m$ denotes the number of items. The attacker crafts a set of malicious user profiles denoted as $\mathcal{U}^a=\{u^a_1,u^a_2, \dots, u^a_k\}$, with $k=|\mathcal{U}^a|$, and generates the corresponding interaction matrix $\mathbf{Y}^a\in \{0,1\}^{k\times m}$ for these fabricated users. This process is formalized as follows:
\begin{equation}
\mathbf{Y}^a = A(\mathbf{Y}, \mathcal{T}, k),
\end{equation}
where $A(\cdot)$ represents the shilling attack algorithm, which takes the existing user-item interactions $\mathbf{Y}$, the set of target items $\mathcal{T}$, and the desired number of fake profiles $k$ as inputs, and produces the manipulated interaction matrix $\mathbf{Y}^a$. Subsequently, the attacker injects these malicious user profiles into the recommendation system, augmenting the original user-item matrix to $\mathbf{Y}'\in \{0,1\}^{(n+k)\times m}$, thus influencing the recommendation outcomes. 
Following previous works~\cite{AUSH, HRGAN, LegUP, GSPAttack}, our method ToDA focus on promotion attacks, which aim to maximize the availability of the target items in the recommendation list, in the current study to demonstrate the effectiveness and efficiency.

\subsubsection{Attacker capability}

The efficacy of shilling attacks is also contingent upon various factors that delineate the attacker's capabilities. These factors are assessed through a practical lens as outlined below:

\begin{itemize}
\item \textbf{Incomplete data:} Shilling attackers are assumed to operate with limited information, reflecting only a subset of the total user-item interactions.
Prior works~\cite{AUSH, HRGAN, LegUP} often leverage the complete interactions of RS, countering the real scenario with system's access constrains.
Therefore, we define that the malicious user profiles, $\mathcal{U}^a$, and corresponding interaction matrix, $\mathbf{Y}^a$, are built upon this incomplete data set.

\item \textbf{Black-box setting:} Attackers do not have access to the inner workings of the recommendation system, including its learning algorithm and parameter settings. The shilling attack algorithm, $A(\cdot)$, therefore, operates blindly, producing $\mathbf{Y}^a$ without detailed insights about the system. There is a general solution that introduce surrogate model trained with gleaned data to represent the victim model~\cite{LegUP, GSPAttack}. However, such method is extensively time-consuming since it requires continuously updating the surrogate model.
In contrast, ToDA can optimize the attack objective without surrogate model-aided and achieves outperformed efficacy than SOTA.

\item \textbf{No extra knowledge:} Attacker doesn't possess additional knowledge outside of the gleaned interactions, $\mathbf{Y}'$. This includes knowledge about item features, user demographics, or other auxiliary information that could enhance the attack's precision. 
Due to the incomplete data, several methods turn to introduce cross-domain data or knowledge graph to enhance the attack capability~\cite{CopyAttack, PCAttack}, then boosting the attack performance, or exploring on domain-specific recommendations (like review-based~\cite{ARG} or sequential recommendation~\cite{seqAttack}).
However, we aim to conduct the practical attack without extra knowledge against general recommendation system, then fundamentally facilitating the development of shilling attacks.

\end{itemize}


\subsection{Recommendation System}

Let $\mathcal{U} = \{u_1, u_2, \dots, u_n\}$ represent the set of users and $\mathcal{I} = \{i_1, i_2, \dots, i_m\}$ denote the set of items. The implicit interactions between users and items are encoded in a binary matrix $\tilde{\mathbf{Y}} \in \{0, 1\}^{n \times m}$, where $\tilde{\mathbf{Y}}_{u,i}$ equals $1$ if user $u$ has interacted with item $i$, and $0$ otherwise.

The primary objective of recommendation system is to predict the preference score $\hat{\mathbf{Y}}_{u,i}$, estimating the likelihood of user $u$ engaging with item $i$. This preference score is computed by a recommendation model $R$, which leverages collaborative filtering techniques to analyze the historical interactions captured in $\tilde{\mathbf{Y}}$. Mathematically, the recommendation model is represented as:
\begin{equation}
\hat{\mathbf{Y}}_{u,i} = R(u, i, \tilde{\mathbf{Y}}),
\end{equation}
where $R(\cdot)$ utilizes the observed user-item interactions to infer the preference of user $u$ towards item $i$. Through this process, the recommendation system aims to enhance user satisfaction and engagement by delivering relevant and personalized recommendations.

%% file: sections/4-Method.tex
\section{Methodology}

\begin{figure*}
    \centering
    \includegraphics[width=\textwidth]{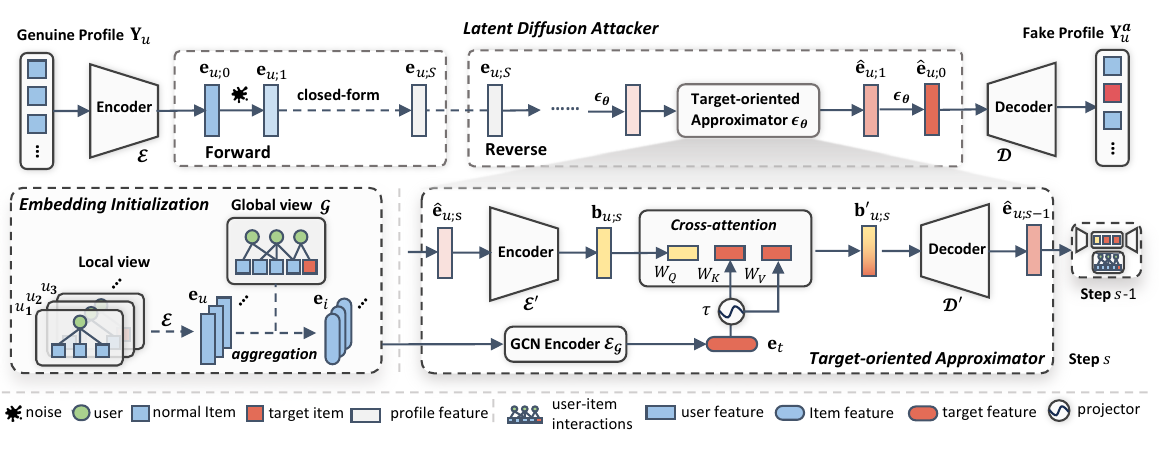}
    \vspace{-8mm}
    \caption{The overall framework of ToDA. It is featured by the latent diffusion attacker with target-oriented approximator $\boldsymbol{\epsilon}_\theta$. The approximator $\boldsymbol{\epsilon}_\theta$ denoises the previous state of user profile feature $\mathbf{e}_{u;s}$, bringing in the global information to the local view diffusion model. The features of target items are incorporated through a GCN Encoder $\mathcal{E}_{\mathcal{G}}$, before using the cross-attention to control the denoising orientation. }
    \label{fig:main_framework}
    \vspace{-4mm}
\end{figure*}

We introduce the overall framework of our proposed method (in Section~\ref{sec:overview}), and specifically elucidate the crux of ToDA (as shown in Figure~\ref{fig:main_framework}), the Latent Diffusion Attacker (in Section~\ref{sec:LDA}), which adds noise to latent features and reverse them through Target-oriented Approximator (in Section~\ref{sec:ToA}), and the optimization of ToDA (in Section~\ref{sec:Opt}). Moreover, the complexity analysis showcases the efficiency of proposed method (in Section~\ref{sec:complexity_analysis}).

\subsection{Approach Overview}
\label{sec:overview}

Our goal is to introduce new paradigm of utilizing diffusion models to generate the user profile, achieving effective shilling attacks.
Such paradigm avoids the previous learning dilemmas and inflexibility, opening a new door for recent still researches of shilling attacks.
To this end, we point out several inevitable challenges, and establish a set of design principles that guide the development of our model:
\begin{itemize}
    \item ~\textit{Attack-Effectiveness (P1)}: A pivotal criterion in our framework is the effectiveness of the shilling attack simulation. This involves the generation of fabricated user profiles designed to manipulate the recommendation systems. 
    To endow the malicious intent to conventional DMs, we aim to construct beneficial target item features, serve as the condition to guide the  generation of DMs. 
    \item ~\textit{Generation-Efficiency (P2)}: Given the dynamic nature of recommendation systems and the computational intensity associated with conventional DMs, we recognize the need for a design that prioritizes efficiency in the generation process. Our approach incorporates sophisticated mechanisms to streamline the generation of synthetic profiles, thereby reducing the computational overhead and facilitating the scalability of both shilling attack simulations and DM applications.
    \item ~\textit{Imperceptibility (P3)}: The imperceptibility of synthetic user profiles is crucial for the efficacy of shilling attacks. However, in the context of our research, this is considered a secondary objective. We acknowledge that while the ability to create imperceptible profiles is important, it is not the primary focus of our study. Nonetheless, the high-fidelity simulation capabilities inherent in DMs and GAN-based methods naturally contribute to the generation of profiles that are difficult to distinguish from genuine user data.
\end{itemize}

In alignment with the aforementioned principles, we introduce novel designs that not only enhance their utility in the context of shilling attack tasks but also address the limitations of conventional DMs:
\begin{itemize}
    \item ~\textit{Global Target-oriented Approximator (D1)}: To fulfill the Attack-Effectiveness \textbf{(P1)}, we augment DMs with the capacity to simulate shilling attacks by integrating conditional generative frameworks. Our approach transcends the traditional localized interactions of DMs by employing a global interaction graph. This graph enables the generation of a comprehensive and informative target item representation that guides the generative process with distinct and robust signals, ensuring a higher fidelity in simulating targeted shilling attacks.
    \item ~\textit{Lightweight Architecture (D2)}: Acknowledging the computational complexity of DMs, we propose a streamlined architecture \textbf{(P2)} for the dynamics of recommendation system. Our design eliminates redundant and cumbersome components, opting instead for a more efficient latent process that significantly reduces computational demands. Furthermore, we have replaced the traditional U-Net architecture with our proposed Target-oriented Approximator (ToA), as detailed in (D1). This novel architecture retains the core conceptual framework of DMs while simplifying complex and repetitive modules, leading to a more lightweight and agile model.
\end{itemize}

To integrate design principles with diffusion models, we introduce the Latent Diffusion Attacker (LDA), which is innated  with a high-fidelity simulation capability \textbf{(P3)}. 
Moreover, the LDA transcends the mere application of pre-existing diffusion models by incorporating a lightweight latent diffusion process, being complemented by a sophisticated mechanism for guiding the construction of user profiles, tailored to target items.

\subsection{Latent Diffusion Attacker}
\label{sec:LDA}

DMs are employed as the foundational attacker model for user profile generation in our approach. 
In a typical DMs' process, noise is gradually added at each forward step, while the reverse operation attempts to reconstruct the state from the preceding step.
These processes are guided by the Markov assumption, implying that each state is exclusively reliant on its immediate predecessor.
Consequently, these two processes can be formally represented as follows: 

\hspace{-4mm}
\textbf{Forward:}
\begin{equation}
    q(\mathbf{Y}_{u; 1:S}|\mathbf{Y}_{u;0})=\prod_{s=1}^{S} q(\mathbf{Y}_{u;s} | \mathbf{Y}_{u; s-1}),
    \label{eq:forward_diffusion}
\end{equation}
where $q(\mathbf{Y}_{u; 1:S}|\mathbf{Y}_{u;0})$ represents the probability distribution of the sequence of user profiles during the forward procedure, which includes $S$ steps. The product operation indicates that at each step $s$, the next profile $\mathbf{Y}_{u;s}$ is conditionally dependent only on the preceding one $\mathbf{Y}_{u; s-1}$.

\hspace{-4mm}
\textbf{Reverse:}
\begin{equation}
    p(\mathbf{Y}_{u; 0:S}) = p(\mathbf{Y}_{u; 0})\prod_{s=1}^S p(\mathbf{Y}_{u;s-1}|\mathbf{Y}_{u;s}),
    \label{eq:reverse_diffusion}
\end{equation}
where $p(\mathbf{Y}_{u; 0:S})$ denotes the probability distribution of the sequence of user profiles during the reverse process. The term $p(\mathbf{Y}_{u; 0})$ stands for the probability of the initial profile (\ie $\mathbf{Y}_{u; 0} = \mathbf{Y}_{u}$), while the product term represents the conditional probability of each preceding profile $\mathbf{Y}_{u;s-1}$ given its successor $\mathbf{Y}_{u;s}$, iteratively applied over all $S$ steps.
Following these procedure, The conventional DMs achieve reconstruction of the user profiles in a discrete state, to ensure the imperceptibility.  
However, such discrete state has a large process space which aligns with the size of items, thus increasing the computational burden and further hindering the flexibility of conditional generation for DMs.

To facilitate the malicious profile generation and diminish the computational overhead during training, we adopt principles akin to perceptual image compression in computer vision~\cite{LDM}, exploiting the benefits of high-dimensional spaces.
More precisely, given a user profile $\mathbf{Y}_{u}\in\{0,1\}^{m}$, the encoder $\mathcal{E}$ encodes $\mathbf{Y}_{u}$ into a latent representation $\mathbf{e}_u = \mathcal{E}(\mathbf{Y}_{u}) \in \mathbb{R}^{d}$ and the decoder $\mathcal{D}$ reconstructs the user profile $\mathbf{Y}^a_{u} = \mathcal{D}(\mathbf{e}_u)$. 
Wherein, both the encoder $\mathcal{E}$ and decoder $\mathcal{D}$ are implemented by the multilayer perceptron (MLP), following the architecture of MultiDAE~\cite{multivae}.
And we calculate the multinomial likelihood to optimize the parameters of $\mathcal{D}$ and $\mathcal{E}$, formally,
\begin{equation} 
    \mathcal{L}_r  = -\mathbb{E}_{i\in \mathcal{I}} \mathbf{Y}_{u,i}\log{\mathcal{D}(\mathbf{e}_u)}.
    \label{eq:reconstruct}
\end{equation}
After the training via $\mathcal{L}_r$, we obtain the paired autoencoder that is capable to transform discrete user profiles into continual user features. 
Thereby we construct the latent diffusion attacker inheriting the Equation~\ref{eq:forward_diffusion} and~\ref{eq:reverse_diffusion} as
$q(\mathbf{e}_{u; 1:S}|\mathbf{e}_{u; 0})$ and $p(\mathbf{e}_{u; 0:S})$, respectively.

For the forward process, we add noise to the user feature step by step following: 
\begin{equation}
    q(\mathbf{e}_{u;s} \vert \mathbf{e}_{u;s-1}) = \mathcal{N}(\mathbf{e}_{u;s}; \sqrt{1 - \beta_s} \mathbf{e}_{u;s-1}, \beta_s\mathbf{I}),
\end{equation}
where $\{\beta_s\}^S_{s=1}$ is the variance schedule to control the noise scale. 
And we exploit a reparameterization trick ~\cite{DDPM} to formulate the distribution after adding the gaussian noise. 
Such trick also allows for a closed-form solution for the forward procedure, which means that we can directly compute the user features at a future step $s$, $\mathbf{e}_{u;s}$, from the initial state $\mathbf{e}_{u;0}$: 
\begin{equation}
q(\mathbf{e}_{u;s} \vert \mathbf{e}_{u;0}) = \mathcal{N}(\mathbf{e}_{u;s}; \sqrt{\bar{\alpha}_t} \mathbf{e}_{u;0}, (1 - \bar{\alpha}_s)\mathbf{I}),
\end{equation}
where $\bar{\alpha}_s = \prod_{i=1}^s (1-\beta_i)$.
For the reverse process, we aim to iteratively denoise $\mathbf{e}_{u;s}$ for $s$ steps to obtain the initial user feature $\mathbf{e}_{u;0}$:
\begin{equation}
    p(\mathbf{e}_{u;s-1}|\mathbf{e}_{u;s}) = \mathcal{N}(\mathbf{e}_{u;s}; \boldsymbol{\epsilon}_\theta(\mathbf{e}_{u;s}, s, \mathbf{e}_{t}), \Sigma_\theta(\mathbf{e}_{u;s}, s)),
\end{equation}
where $\Sigma_\theta(\mathbf{e}_{u;s}, s)=\frac{1-\bar{\alpha}_{s-1}}{1-\bar{\alpha}_{s}}\beta_s\mathbf{I}$ denotes the variance and the target-oriented approximator $\boldsymbol{\epsilon}_\theta(\cdot)$ takes the target item feature $\mathbf{e}_{t}$ as the conditional signal to guide the reversion procedure.
To optimize the latent diffusion model, DMs compel the posterior distribution closer to the prior distribution via KL divergence:
\begin{equation}
    \begin{aligned}
    \mathcal{L}_d
    &= \mathbb{E}_{s \sim [1, S], \mathbf{e}_{u;0}, \boldsymbol{\epsilon}_s} \text{KL}(q(\mathbf{e}_{u;s-1}|\mathbf{e}_{u;0}) || p(\mathbf{e}_{u;s-1}|\mathbf{e}_{u;s}))
    \end{aligned}
    \label{eq:diff_loss}
\end{equation}
Thanks to the DDPM framework~\cite{DDPM}, the above objective can be simplified to a Mean Squared Error (MSE) between the noise $\boldsymbol{\epsilon}_s$ and the estimated one approximated by $\boldsymbol{\epsilon}_\theta$ as follows:
\begin{equation}
    \begin{aligned}
    \mathcal{L}_d
    &= \mathbb{E}_{s \sim [1, S], \mathbf{e}_{u;0}, \boldsymbol{\epsilon}_s} \Big[\|\boldsymbol{\epsilon}_s - \boldsymbol{\epsilon}_\theta(\mathbf{e}_{u;s}, s, \mathbf{e}_{t})\|^2 \Big] \\
    &= \mathbb{E}_{s \sim [1, S], \mathbf{e}_{u;0}, \boldsymbol{\epsilon}_s} \Big[\|\boldsymbol{\epsilon}_s - \boldsymbol{\epsilon}_\theta(\sqrt{\bar{\alpha}_s}\mathbf{e}_{u;0} + \sqrt{1 - \bar{\alpha}_s}\boldsymbol{\epsilon}_s, s, \mathbf{e}_{t})\|^2 \Big].
    \end{aligned}
    \label{eq:diff_loss}
\end{equation}

\subsection{Target-oriented Approximator}
\label{sec:ToA}

The latent diffusion model achieves efficient generation of user profiles, while lacking the malicious intent.
To endow the diffusion attacker with the target-oriented ability, we tailor the estimator $\epsilon_\theta$ with a global horizon view to compensate the narrow focus of DMs and leverage the cross attention to transfer the target information into the latent user feature. 

To obtain the global view item feature, we commerce with using the encoded user profile features $\mathbf{e}_{u}$ to initialize the item embeddings as:
\begin{equation}
    \mathbf{e}_i = \frac{1}{|\mathcal{N}_i|}\sum_{u\in \mathcal{N}_i} \mathbf{e}_u,
    \label{eq:item_embedding}
\end{equation}
where $\mathcal{N}_i$ is the indices of users that are interacted with item $i$.
By aggregating the interacted user features, we maintain a consistent space alignment with the latent diffusion attacker, facilitating the space alignment and following transferring.
Inspired by the high-order information capturing, we apply a GNN model to learn the item features from the message-passing mechanism. 
The GNN model is adept at processing graph-structured data, thereby enabling the extraction of informative representations for nodes, where interactions among users and items can be naturally modeled as a graph, denoted as $\mathcal{G} = (\{\mathcal{U}, \mathcal{I}\}, \mathbf{Y})$.
$\{\mathcal{U}, \mathcal{I}\}$ denotes the nodes while $\mathbf{Y}$ represents the edges of graph $\mathcal{G}$.
Obviously, the integration of graphs extends the generation of a broader horizon outwards the on-processing user feature itself, compensating the narrow focus of DMs. 
Formally, we learn the target item features as: 
\begin{equation}
    \mathbf{e}_t = \mathcal{E}_{\mathcal{G}}(t),\quad
    \mathcal{G} = (\mathbf{e}_{\mathcal{U}}, \mathbf{e}_{\mathcal{I}},\mathbf{Y}),
\end{equation}
where $\mathcal{E}_{\mathcal{G}}$ is the GNN encoder, which is equipped with the user-item bipartite graph $\mathcal{G}$, and takes the index of the target item $t$ as input. 
$\mathbf{e}_{\mathcal{I}}$ is the matrix consisting of all item embeddings generated by Equation ~\ref{eq:item_embedding}.
Specifically,
we remove the parameters of conventional GCN to achieve efficient feature aggregation via normalized symmetric adjacency matrix: $\hat{\mathbf{A}} = \mathbf{D}^{-1/2}\mathbf{A}\mathbf{D}^{-1/2}$~\cite{LightGCN}, therefore seamlessly obtaining the target item features from the global structure of user-item interactions. 

In the context of DMs, an approximator is employed to estimate noise and progressively refines the input $\mathbf{e}_{u;S}$.
While this method is exquisite, it overlooks the significance of target items.
Its primary focus remains ensuring that shilling attacks go unnoticed.
To rectify this limitation and bolster the efficacy of the attacks, we enhance the original approximator, denoted as $\epsilon_\theta$, to focus on the target items, reaping the benefits of both efficacy and efficiency. 
This modified approximate takes into consideration the user's latent feature $\mathbf{\hat{e}}_{u;s}$ at the preceding step $s$ and the target item feature $\mathbf{e}_{t}$, then producing the next user latent feature $\mathbf{\hat{e}}_{u;s-1}$.
For the purpose of efficiently handling the information related to target items, we employ AE models, which allows us to compress the original feature with $\mathcal{E}'$ into a latent bottleneck $\mathbf{b}_{u;s}$, and then to reconstruct it using $\mathcal{D}'$. 
To pre-process $\mathbf{e}_t$, we introduce a transformation $\tau$ that projects $\mathbf{e}_t$ to $\tau(\mathbf{e}_t)$ aligned with the $\mathbf{b}_{u;s}$, which is then carried out through a cross-attention layer implementing 
\begin{equation}
\begin{aligned}
    & Attn(\mathbf{Q},\mathbf{K},\mathbf{V})=\text{softmax}(\frac{\mathbf{Q}\mathbf{K}^T}{\sqrt{d}})\cdot \mathbf{V},\\
    \mathbf{Q} =  & \mathbf{W}_Q\cdot \mathbf{b}_{u;s},\   
    \mathbf{K} = \mathbf{W}_K\cdot \tau(\mathbf{e}_{t}),\  
    \mathbf{V} = \mathbf{W}_V\cdot \tau(\mathbf{e}_{t}),
\end{aligned}
\end{equation}
where $\mathbf{W}_Q, \mathbf{W}_K, \mathbf{W}_V\in \mathbb{R}^{d'\times d}$ are learnable matrices. Note that, our method is also flexible to incorporate multiple target items, which can be interpreted in a matrix format:  $\mathbf{Q} \in \mathbb{R}^{|\mathcal{U}'|\times d'}$, and $\mathbf{K}, \mathbf{V}\in \mathbb{R}^{|\mathcal{T}| \times d'}$, where $|\mathcal{U}'|$ represents a batch of users and $|\mathcal{T}|$ denotes the number of target items. Therefore, $\mathbf{Q} \times \mathbf{K}^T \in \mathbb{R}^{|\mathcal{U}'|\times |\mathcal{T}|}$, leading to the final representation $Attn(\mathbf{Q},\mathbf{K},\mathbf{V}) \in \mathbb{R}^{|\mathcal{U}'|\times d'}$, which is guided by multiple target items from $\mathcal{T}$.
A visual representation of the denoising step $s$ is provided in Figure~\ref{fig:main_framework}.
By performing $S$ steps of this target-oriented noise approximation, we obtain the fake user profile $\mathbf{Y}^a_u$ at the end which reaps the benefits of both imperceptibility and attack-effectiveness. 
Notably, conventional DMs typically employ complicated and heavy modules such as U-Net~\cite{u_net} as approximators, constrained by significant computational demands. In contrast, our method is notably more efficient, balancing lightweight design with effectiveness. Further analysis on this is detailed in Section~\ref{sec:complexity_analysis}.

\subsection{Optimization}
\label{sec:Opt}

We implement a hierarchical optimization strategy to update the parameters of ToDA. 
Specifically, we pre-train the initial encoder $\mathcal{E}$ and decoder $\mathcal{D}$, which serve as transforming discreate user profiles into a latent space, by optimizing $\mathcal{L}_{r}$.
Then the latent diffusion attacker is optimized by the defined $\mathcal{L}_d$.
During the inference phase, genuine user profiles are utilized as templates. 
Initially, these templates are transformed via pre-trained encoder $\mathcal{E}$, represented as $\mathbf{e}_u$ for user $u$. 
And noise is added to it in a closed-form solution.
At each step of the reversion process, Gaussian noise is sampled and employed alongside $\boldsymbol{\epsilon}_\theta$ to ascertain the preceding state of user features. 
Following $S$ steps of reversion, the crafted user profile is obtained using a pre-trained decoder $\mathcal{D}$.
And the details of the training and inference are presented in the supplemental materials.

\subsection{Complexity Analysis}
\label{sec:complexity_analysis}
ToDA utilizes the hierarchical training, the first step of which involves training the MLP (Multilayer perceptron)-based autoencoder. The complexity of the autoencoder is $O(\mu d)$, where $\mu$ represents the length of profiles and $d$ is the dimension of the embedded features. Subsequently, we train the main component of ToA, whose complexity is $O(d^2 + d d'^2)$. Here, $d'$ is the dimensionality of the query, key, and value matrix in the cross-attention. The complexity of GNN encoder is $O(\vert\mathbf{E}\vert^2 d)$, where $\vert\mathbf{E}\vert^2 d$ is the number of edges in the user-item graphs. Roughly, the overall complexity is $O(\mu d + \vert\mathbf{E}\vert^2 d + d^2 + d d'^2)$. Conventional DMs often employ U-Net~\cite{u_net}, which is more time-consuming due to the CNN model. In contrast, we tailor the approximator to be lighter, resulting in reduced training time. GSPAttack~\cite{GSPAttack}, the SOTA method for shilling attacks, uses GAN (MLP-based generator and discriminator) as the main architecture to generate profiles, equipped with a GNN-based surrogate model. Thus, the complexity of GSPAttack is $O((\mu d + d^2) \times \vert\mathbf{E}\vert^2 d)$, cosuming much time than ours. For another method AUSH~\cite{AUSH}, it only uses MLP-based generator and discriminator and has smaller complexity of $O(\mu d)$. However, its performance is worse than ours with a noticeable margin. 
See the empirical analysis in the Sec.~\ref{sec:practicality}.
In conclusion, through the comparison with conventional DMs and these two methods, ToDA alleviates the computation costs of DMs and exhibits efficient and effective performance.


%% file: sections/5-Experiment.tex
\section{Experiment}

\begin{table}
\centering
\caption{The statistics of the datasets.}

\label{tab:data_statistics}
\renewcommand{\arraystretch}{1}
\setlength{\tabcolsep}{2mm}
\begin{tabular}{lcccc} 
\hline
\textbf{Dataset} & \textbf{\#user} & \textbf{\#item} & \textbf{\#inter.} & \textbf{sparsity}  \\ 
\hline\hline
ML-100K          & 943             & 1682            & 100,000           & 93.70\%      \\
FilmTrust        & 1508            & 2071            & 35,494            & 98.86\%            \\
Gowalla          & 29,858          & 40,981          & 1,027,370         & 99.92\%        \\
\hline
\end{tabular}
\vspace{-4mm}
\end{table}

We conduct extensive experiments to demonstrate the effectiveness of ToDA, and perform in-depth analysis by addressing the following research questions (RQs):
\begin{itemize}
    \item{\textbf{RQ1}}: Does ToDA outperform the state-of-art models for shilling attack, including heuristic-based, optimization-based, and generative-based methods?
    \item{\textbf{RQ2}}: How does the shilling attack efficacy benefit from each component of ToDA, including the target-oriented approximator and its cross-attention layer?
    \item{\textbf{RQ3}}: Does ToDA ensure the praticality and the imperceptibility?
\end{itemize}

To provide a comprehensive elucidation of these research questions, we adopt the following structured approach in this section. First, we detail the experimental settings in Sec.~\ref{sec:experimental_setting}, including datasets, evaluation metrics and selected baselines. Next, Sec.~\ref{sec:overall_performance} provides a comparison of performance between our proposed model and competitive shilling attack models across all three widely adopted public datasets. Furthermore, to ensure the fairness and accuracy of the experimental results, we select three representative models as the victim recommendation systems (\ie MF~\cite{BPR}, LGN~\cite{LightGCN} and NCF~\cite{NCF}).
Following this, in Section~\ref{sec:ablation_study}, we conduct ablation studies to investigate the specific impact of each component on attack efficiency.
At the end, we explore the practicality of ToDA, demonstrate the imperceptibility of the generated user profiles, and further analyze the impacts on different hyper-parameter settings (\ie the diffusion step $S$, the noise scale $\beta$, learning rate and L2 regularization) in Sec.~\ref{sec:model_study}.

\subsection{Experimental Settings}
\label{sec:experimental_setting}
We introduce the settings of experiments, including datasets for evaluation, metrics, baselines and different hyper-parameters.

\subsubsection{Datasets}

Following previous shilling attack studies\cite{AUSH, LegUP}, we employ three widely used public datasets for experimental evaluation, encompassing ML-100K\footnote{https://grouplens.org/datasets/movielens/100k/}, FilmTrust\footnote{https://www.librec.net/datasets/filmtrust.zip}, and Gowalla\footnote{https://www.gowalla.com}. The statistics of these datasets are detailed in Table \ref{tab:data_statistics}. These datasets exhibit distinct characteristics in terms of user and item cardinality, as well as the density of their interaction matrices.
Wherein, ML-100K and FilmTrust are commonly utilized in shilling attack tasks. 
To align closely with the context of RS, we incorporate Gowalla, a dataset widely used in RS research. 
Gowalla has a substantially larger user and item base, providing a more comprehensive dataset for our analysis.
To better simulate real-world recommender systems and attackers, we utilize the entire dataset to train the victim model, while randomly sampling 25\% of the data from the dataset exclusively for training the attack models. For each dataset, we use the ratio 8:1:1 to randomly split the historical interactions and constituted the training set and testing set for the victim models, following the standard training and evaluation strategy of recommendation systems~\cite{LightGCN}.

\subsubsection{Evaluation Metrics}
For performance comparison, we adopt two widely used metrics Average Hit Ratio (HR@\textit{K}) and Mean Reciprocal Rank (MRR@\textit{K}), where K is set to 10 to limit the length of the recommended item list. 
The HR@\textit{K} metric measures the proportion of times that at least one relevant item is recommended within the top-k items for a user, while the MRR@\textit{K} calculates the average of the reciprocal ranks of the first relevant item in the recommendation list for each user. 
Higher HR@\textit{K} value and MRR@\textit{K} value indicate that the system is more successful at suggesting relevant items. 
Additionally, we introduce 50 fabricated user profiles into the dataset to obtain the polluted datasets, then take it for training the victim model.
This aligns with the settings established in previous relevant research \cite{AUSH, LegUP}.

\subsubsection{Baselines}
To evaluate the boosting of our proposed method, we compare it with several SOTA methods. We briefly divide these baselines into three groups: 1) heuristic-based models (\ie Random Attack, Average Attack, and Bandwagon Attack), 2) optimization-based methods (\ie DL-Attack), and 3) generative-based approaches (\ie AUSH, LegUP, and GSPAttack) as follows:
\begin{itemize}
    \item \textbf{Random Attack}~\cite{shillingAttack} generates fake user profiles using a stochastic item selection mechanism, with the additional step of specifying target items within these artificial profiles, thereby generating random user-item interactions.
    \item \textbf{Average Attack}~\cite{shillingAttack} 
    crafts fabricated user profiles by leveraging global statistics extracted from the victim system. Compared with Random Attack, this method focuses on item selection according to their frequency of occurrence.
    \item  \textbf{Bandwagon Attack}~\cite{bandwagon} 
    leverages popularity bias in recommendation systems to enhance attack efficiency. Different from Average Attack, it generates fake user profiles based on the most frequently interacted items.
    \item \textbf{DLAttack}~\cite{DLAttack}~transforms the complex optimization problem with approximations to derive a loss function and then iteratively trains a "poison" model.
    \item \textbf{AUSH}~\cite{AUSH} utilizes GAN techniques to create fake user profiles through the generator module. And the discriminator ensures the imperceptibility of user profiles.
    \item \textbf{LegUP}~\cite{LegUP} extends AUSH by introducing an additional surrogate  model. A designed loss function is also adopted to strengthen the attack effectiveness of the model.
    \item \textbf{GSPAttack}~\cite{GSPAttack} is a SOTA method for shilling attack in recommendation system. This method uses a GNN as a surrogate model, which is then fused to generate the fabricated profile.
\end{itemize}

\subsubsection{Hyper-parameter Settings}
We set the latent dimension to 64 and adhere to corresponding articles to achieve optimal performance for the victim models. 
For weight initialization, the Xavier Initialization~\cite{xavier} is employed, while optimization is carried out using the Adam Optimizer~\cite{Adam}.
To find the optimal hyper-parameter setting, we adopt the grid search strategy.
Specifically, we adjust the learning rate in range of $\{10^{-2}, 10^{-3}, 10^{-4}\}$, and weight decay in range of $\{10^{-3}, 10^{-4}, 10^{-5}\}$. 
The $S$ and the number of GNN layers are tuned within $\{1, 10, 50, 100, 500, 1000\}$ and $\{1,2,3\}$, respectively. 
To exhibit the influence of noise comprehensively, we adjust $\beta$ in ranges of $\{[10^{-4}, 2\times10^{-4}], [10^{-4}, 10^{-3}], $ $[10^{-3}, 2\times10^{-3}], [10^{-3}, 10^{-2}],[10^{-2},2\times10^{-2}], [10^{-2}, 10^{-1}]\}$.
We involve targeting three widely acknowledged recommendation systems: 
\begin{itemize}
    \item \textbf{MF}~\cite{BPR} primarily operates by learning latent embeddings of both users and items to subsequently calculate ranking scores based solely on these embeddings.
    \item \textbf{NCF}~\cite{NCF} uses a Multi-Layer Perceptron (MLP) to learn user interactions, bringing deep learning to collaborative filtering.
    \item \textbf{LGN} (\ie LightGCN)~\cite{LightGCN}, on the other hand, leverages the graph structure information and employs Graph Convolutional Network (GCN) techniques to attain more robust user and item representations.
\end{itemize}
These algorithms form the basis for many modern recommendation systems~\cite{recomeSurvey, GNNRec_survey}.
The same settings are employed to implement and evaluate the baselines to ensure a fair comparison.

\subsection{Performance Comparison (RQ1)}
\label{sec:overall_performance}

\begin{table*}
\renewcommand{\arraystretch}{1}
\centering
\caption{The overall performance comparison, where the strongest baselines are underlined and the best results are bold.}
\label{tab:overall_comparison}
\begin{tabular}{l|cc|ccc|c|ccc|c} 
\hline
\textbf{Datasets}          & \textbf{Victim}      & \textbf{Metric} & \textbf{\textbf{Random~}} & \textbf{\textbf{Average~}} & \textbf{\textbf{Bandwagon~}} & \textbf{DLAttack} & \textbf{\textbf{AUSH}} & \textbf{\textbf{LegUP}}            & \textbf{GSPAttack} & {\cellcolor[rgb]{0.902,0.902,0.902}}\textbf{ToDA}                        \\ 
\hline\hline
\multirow{6}{*}{ML-100K}   & \multirow{2}{*}{MF}  & HR              & 0.0711                    & 0.0785                     & 0.0747                       & 0.0946            & 0.0903                 & 0.0819                             & \uline{0.1043}     & {\cellcolor[rgb]{0.902,0.902,0.902}}\textbf{0.1156}                      \\
                           &                      & MRR             & 0.0180                    & 0.0210                     & 0.0175                       & 0.0292            & 0.0249                 & 0.0228                             & \uline{0.0322}     & {\cellcolor[rgb]{0.902,0.902,0.902}}\textbf{0.0407}                      \\ 
\hhline{~----------}
                           & \multirow{2}{*}{LGN} & HR              & 0.0823                    & 0.0795                     & 0.0855                       & 0.0853            & \uline{0.0914}         & 0.0876                             & 0.0882             & {\cellcolor[rgb]{0.902,0.902,0.902}}\textbf{0.0971}                      \\
                           &                      & MRR             & 0.0232                    & 0.0226                     & 0.0244                       & 0.0268            & \uline{0.0266}         & 0.0274                             & 0.0259             & {\cellcolor[rgb]{0.902,0.902,0.902}}\textbf{0.0307}                      \\ 
\hhline{~----------}
                           & \multirow{2}{*}{NCF} & HR              & 0.0708                    & 0.0766                     & 0.0861                       & 0.0817            & 0.0672                 & \multicolumn{1}{l}{\uline{0.0946}} & 0.0838             & \multicolumn{1}{l}{{\cellcolor[rgb]{0.902,0.902,0.902}}\textbf{0.1016}}  \\
                           &                      & MRR             & 0.0150                    & 0.0158                     & 0.0167                       & 0.0155            & 0.0138                 & \multicolumn{1}{l}{\uline{0.0189}} & 0.0168             & \multicolumn{1}{l}{{\cellcolor[rgb]{0.902,0.902,0.902}}\textbf{0.0211}}  \\ 
\hline
\multirow{6}{*}{FilmTrust} & \multirow{2}{*}{MF}  & HR              & 0.1899                    & 0.1931                     & 0.1935                       & 0.1939            & 0.1942                 & \uline{0.1943}                     & 0.1918             & {\cellcolor[rgb]{0.902,0.902,0.902}}\textbf{0.1966}                      \\
                           &                      & MRR             & 0.0854                    & 0.0940                     & 0.0895                       & 0.0939            & \uline{0.0945}         & 0.0936                             & 0.0849             & {\cellcolor[rgb]{0.902,0.902,0.902}}\textbf{0.0974}                      \\ 
\hhline{~----------}
                           & \multirow{2}{*}{LGN} & HR              & 0.1826                    & 0.1924                     & 0.1930                       & 0.1909            & 0.1963                 & 0.1924                             & \uline{0.1976}     & {\cellcolor[rgb]{0.902,0.902,0.902}}\textbf{0.1983}                      \\
                           &                      & MRR             & 0.0898                    & 0.0936                     & 0.0943                       & 0.0905            & 0.0960                 & 0.0942                             & \uline{0.0972}     & {\cellcolor[rgb]{0.902,0.902,0.902}}\textbf{0.0990}                      \\ 
\hhline{~----------}
                           & \multirow{2}{*}{NCF} & HR              & 0.1895                    & 0.1890                     & 0.1902                       & \uline{0.1983}    & 0.1934                 & \multicolumn{1}{l}{0.1927}         & 0.1939             & \multicolumn{1}{l}{{\cellcolor[rgb]{0.902,0.902,0.902}}\textbf{0.1997}}           \\
                           &                      & MRR             & 0.0192                    & 0.0190                     & 0.0178                       & 0.0189            & \uline{0.0204}         & \multicolumn{1}{l}{0.0183}         & 0.0188             & \multicolumn{1}{l}{{\cellcolor[rgb]{0.902,0.902,0.902}}\textbf{0.0205}}           \\ 
\hline
\multirow{6}{*}{Gowalla}   & \multirow{2}{*}{MF}  & HR              & 0.0188                    & 0.0209                     & 0.0197                       & 0.0212            & 0.0253                 & 0.0231                             & \uline{0.0262}     & {\cellcolor[rgb]{0.902,0.902,0.902}}\textbf{0.0270}                      \\
                           &                      & MRR             & 0.0066                    & 0.0074                     & 0.0067                       & 0.0078            & 0.0094                 & 0.0082                             & \uline{0.0098}     & {\cellcolor[rgb]{0.902,0.902,0.902}}\textbf{0.0111}                      \\ 
\hhline{~----------}
                           & \multirow{2}{*}{LGN} & HR              & 0.0234                    & 0.0240                     & 0.0247                       & 0.0286            & 0.0261                 & 0.0266                             & \uline{0.0275}     & {\cellcolor[rgb]{0.902,0.902,0.902}}\textbf{0.0294}                      \\
                           &                      & MRR             & 0.0082                    & 0.0086                     & 0.0095                       & 0.0116            & 0.0099                 & 0.0104                             & \uline{0.0105}     & {\cellcolor[rgb]{0.902,0.902,0.902}}\textbf{0.0120}                      \\ 
\hhline{~----------}
                           & \multirow{2}{*}{NCF} & HR              & 0.0396                    & 0.0432                     & 0.0408                       & 0.0360            & 0.0495                 & \multicolumn{1}{l}{0.0375}         & \uline{0.0501}     & \multicolumn{1}{l}{{\cellcolor[rgb]{0.902,0.902,0.902}}\textbf{0.0562}}  \\
                           &                      & MRR             & 0.0127                    & 0.0138                     & 0.0121                       & 0.0104            & 0.0147                 & \multicolumn{1}{l}{0.0102}         & \uline{0.0148}     & \multicolumn{1}{l}{{\cellcolor[rgb]{0.902,0.902,0.902}}\textbf{0.0182}}  \\
\hline
\end{tabular}
\vspace{-4mm}
\end{table*}

As detailed in Table \ref{tab:overall_comparison}, we provide an overall performance comparison among baselines and ToDA across three datasets and three victim models. This comparison thoroughly addresses RQ1. Our key observations include the following:
\begin{itemize}
    \item By conducting a comparative analysis, we observe that our proposed model outperforms baselines, within the framework of three victim recommenders. These notable performance improvements become most apparent under the ML-100K and Gowalla datasets, with up to 12.2\% (HR) and 23.0\% (MRR) relative improvements on Gowalla, and 10.8\% (HR) and 26.4\% (MRR) on ML-100K. 
    It is substantiated that the successful integration of attack capabilities into the diffusion model reaffirms the efficacy of ToDA.
    \item All models perform similarly on the FilmTrust dataset. We attribute it to the high sparsity of the interaction matrix in FilmTrust, with a smaller number of users and items. When generating fake user profiles based on historical interactions, the attack model struggles to accurately capture user behavior patterns. 
    Nonetheless, ToDA outperforms other baselines as well on this dataset. We argue that the sophisticated adaption of DMs alleviates the sparse and noisy issue, therefore boosting the performance.
    \item Generally speaking, GSPAttack model slightly outperforms all the baselines, benefiting from the incorporation of GNN, where it generates fake user-item interactions while maintaining data correlation by the inherent architectural advantage.
    \item DNN-based methods and optimization-based methods outperform heuristic-based methods. This performance gap can be attributed to the superior capacity of effectively approximating the underlying data distribution. The flexibility and adaptability of these methods enable them to capture intricate patterns and relationships within the user interaction, thus improving the performance.
    \item It is worthwhile pointing out that the MRR@10 tends to be lower in small datasets, especially FilmTrust, when attacking NCF. 
    This further suggests that insufficient data would lead instability of shilling attacks. 
    Since NCF is an MLP-based method, it tends to produce consistent preference scores in small datasets, which makes improving the order more challenging.
\end{itemize}

\subsection{Ablation Study (RQ2)}
\label{sec:ablation_study}
To investigate the effectiveness of ToDA, we remove main designs individually of ToDA to examine its performance (\ie components analysis) and compare other target condition methods with cross-attention methods in ToA (\ie condition analysis).

\subsubsection{Components Analysis}
To substantiate RQ2 and demonstrate the effectiveness of the target-oriented approximator, we design the following variants of ToDA:
\begin{itemize}
    \item \textbf{w/o-Diff}: ToDA retains a simplified structure, featuring only the core encoder and decoder components. This specific configuration is attained by setting $S$ to zero.
    \item \textbf{w/o-ToA}: ToDA devolves into an attack model that directly generates fake user profile using a conventional diffusion model. The generated user profiles lack access to the global information of items.
\end{itemize}
Table \ref{tab:components} summarizes the performance of ToDA and its variants. It is encouraging to note that our method achieves the best performance on all the three datasets, thus confirming the broad application of the diffusion model and the effectiveness of target-oriented approximator. On the one hand, the denoising ability of the diffusion model ensures the high quality of the generated fake user profiles. On the other hand, the target-orient approximator endows the diffusion model with global horizon and attack ability. 
We observe that the performance of the simplified ToDA model (\ie \textbf{w/o-Diff}) can be inferior to that of a Random Attack. This highlights the necessity for additional modules in previous works to enhance attack capabilities. It also underscores the superiority of our comprehensive ToDA framework, which incorporates global target information within the diffusion process while simultaneously maintaining both imperceptibility and attack effectiveness.

\begin{table}
\centering
\caption{The performance on different components of ToDA w.r.t. HR@10 and MRR@10.}
\renewcommand{\arraystretch}{1}
\label{tab:components}
\begin{tabular}{l|cc|cc|cc} 
\hline
& \multicolumn{2}{c|}{ML-100K} & \multicolumn{2}{c|}{FilmTrust} & \multicolumn{2}{c}{Gowalla}  \\
& HR & MRR                  & HR & MRR                    & HR & MRR                  \\ 
\hline\hline
\textbf{w/o-Diff}                               & 0.0592     & 0.0112      & 0.1916    & 0.0937                          &0.0194     &0.0068           \\
\textbf{w/o-ToA}                                & 0.0840     & 0.0221      & 0.1927    & 0.0944                        &0.0205          & 0.0072                     \\
\rowcolor[rgb]{0.9,0.9,0.9} \textbf{ToDA} &  0.1156    &  0.0407                     & 0.1966     & 0.0974                     & 0.0270     & 0.0111                      \\
\hline
\end{tabular}
\vspace{-4mm}
\end{table}

\subsubsection{Condition Analysis}
To comprehensively investigate how to transfer target information to latent user features, thereby enhancing the model's attack performance, we compare our cross-attention (\textbf{CA}) with two more different operations: element-wise addition (\textbf{Sum}) and vector concatenation (\textbf{Concat}). 
From Table~\ref{tab:collaborative}, we draw the following observations:
Without any doubt, the performance of \textbf{CA} surpasses that of \textbf{Sum} and \textbf{Concat}, particularly on the Gowalla dataset. This can be attributed to its larger number of users and items, as well as greater sparsity, enabling cross-attention to access more global information. 
Observing the results of \textbf{Sum} and \textbf{Concat}, their performance is relatively consistent across the three datasets but falls short of ToDA's performance. It indicates that simple vector operations, such as summation and concatenation, can not effectively handle the information related to target items. The attention mechanism plays a crucial role in improving the efficacy of the attacks.


\begin{table}
\centering
\caption{The performance on different target condition methods w.r.t. HR@10 and MRR@10.}
\renewcommand{\arraystretch}{1}

\label{tab:collaborative}
\begin{tabular}{l|cc|cc|cc} 
\hline
& \multicolumn{2}{c|}{ML-100K} & \multicolumn{2}{c|}{FilmTrust} & \multicolumn{2}{c}{Gowalla}  \\
& HR & MRR                  & HR & MRR                    & HR & MRR                  \\ 
\hline\hline
\textbf{Sum}    & 0.1037     & 0.0311                      &0.1918      &0.0947                         &0.0199    & 0.0069                            \\
\textbf{Concat} & 0.1020    & 0.0298                      &0.1920     &0.0934                      &0.0200     &0.0070                             \\
\rowcolor[rgb]{0.9,0.9,0.9} \textbf{CA}     & 0.1156    &  0.0407            & 0.1966     & 0.0974                  & 0.0270     & 0.0111                    \\
\hline
\end{tabular}

\end{table}

\subsection{Model Study (RQ3)}
\label{sec:model_study}
We investigate the characteristics of ToDA from multiple perspectives. 
We especially exhibits the practicality of ToDA from the empirical results, demonstrating the aforementioned complexity analysis.
And imperceptibility, the basic property of shilling attacks, is comprehensively illustrated by qualitative and quantitative analysis.
The impact of diverse hyper-parameters is reported in the supplemental materials.


\subsubsection{Practicality Analysis}
\label{sec:practicality}
To mitigate the typically computational demands associated with DMs compared to other generative models like GANs, we have developed a lightweight yet effective approximator (ToA). Additionally, by employing a pre-trained autoencoder to transform profiles into a latent space, we significantly reduce time costs.
We present the running time data in Table~\ref{tab:complexity_analysis}. 
Note that all models are trained using NVIDIA Titan-XP and Titan-V GPUs and we record runing times specifically on the Gowalla dataset to better highlight the differences. 
Importantly, these time costs corroborate our complexity analysis detailed in Section~\ref{sec:complexity_analysis}, illustrating that despite the employment of DMs, ToDA's design demonstrates efficiency, underscoring its potential for wider applications.

\begin{table}
\centering
\caption{The computational complexity comparison of the SOTA methods from empirical and theoretical perspectives.}
\renewcommand{\arraystretch}{1}

\label{tab:complexity_analysis}
\begin{tabular}{l|cc} 
\hline
\textbf{Model} & \textbf{Time cost (s)} & \textbf{Time complexity}                              \\ 
\hline\hline
AUSH           & 15.13                  & $O(\mu d)$                                            \\
GSPAttack      & 109.72                 & $O((\mu d + d^2) \times \vert\mathbf{E}\vert^2 d)$    \\
ToDA           & 35.12                  & $O(\mu d + \vert\mathbf{E}\vert^2 d + d^2 + d d'^2)$  \\
\hline
\end{tabular}
\vspace{-4mm}
\end{table}

\subsubsection{Imperceptibility Analysis}
Despite imperceptibility is the secondary objective in research, we analyze the imperceptibility of ToDA following the conventional methods of attack detection and distribution plotting from both quantitative and qualitative perspectives. 
The former utilizes a pre-built detector to evaluate the precision and recall of detection, while the latter visualizes the distributions of different types of user profile. 
Both analyses show that ToDA has imperceptibility consistent with the SOTA approach.
\begin{table*}
\centering
\caption{The detection results at different datasets and methods.}
\renewcommand{\arraystretch}{1}
\setlength{\tabcolsep}{1.3mm}
\label{tab:detection}
\begin{tabular}{ll|ccc|cccc|c} 
\hline
\textbf{Datasets}          & \textbf{Metrics} & \textbf{Random} & \textbf{Average} & \textbf{Bandwagon} & \textbf{DLAttack} & \textbf{AUSH} & \textbf{LegUP} & \textbf{GSPAttack} & \textbf{ToDA}  \\ 
\hline\hline
\multirow{2}{*}{ML-100K}   & Precision        & 0.0444          & 0.0484           & 0.0470             & 0.0468            & 0.0485        & 0.0488         & 0.0446             & 0.0467         \\
                           & Recall           & 0.5385          & 0.5897           & 0.5500             & 0.5238            & 0.5750        & 0.5476         & 0.5385             & 0.5238         \\ 
\hline
\multirow{2}{*}{FilmTrust} & Precision        & 0.0466          & 0.0444           & 0.0465             & 0.0451            & 0.0424        & 0.0443         & 0.0427             & 0.0426         \\
                           & Recall           & 0.5789          & 0.5676           & 0.5366             & 0.5000            & 0.5405        & 0.5250         & 0.4762             & 0.4762         \\ 
\hline
\multirow{2}{*}{Gowalla}   & Precision        & 0.0485          & 0.0468           & 0.0484             & 0.0463            & 0.0484        &   0.0464             & 0.0448             & 0.0403         \\
                           & Recall           & 0.5610          & 0.5500           & 0.5750             & 0.5366            & 0.5476        &   0.5366             & 0.5250             & 0.5278         \\
\hline
\end{tabular}
\vspace{-4mm}
\end{table*}

\hspace{-5mm}
\textbf{Attack Detection.}
To shed light on the imperceptibility of ToDA, we utilize an unsupervised attack detector~\cite{detect} to identify the fake user profiles. 
Table~\ref{tab:detection} shows the results of three datasets and different methods, including baselines and ToDA \wrt Precision and Recall.
In this context, a lower value in these metrics signifies greater stealth or imperceptibility, thereby evading detection more effectively.
It is clear that heuristic-based methods (\ie Random, Average, and Bandwagon) have relatively higher scores than other methods, indicating that they are more easily detected. 
In contrast, ToDA shows similar or better imperceptibility compared with all baselines.
We attribute it to the reconstruction ability of DMs, enabling ToDA to mimic genuine users.

\begin{figure}
    \centering
    \includegraphics[width=0.99\linewidth]{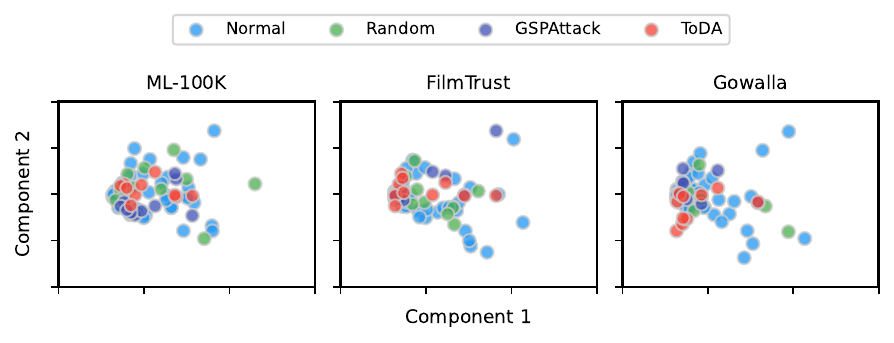}
    \vspace{-4mm}
    \caption{The distributions of user profiles in a 2d-plane.}
    \label{fig:samples}
    \vspace{-4mm}
\end{figure}

\hspace{-5mm}
\textbf{Distribution Plotting.}
To gain an intuitive understanding, we employ Principal Component Analysis (PCA) to project user profiles into a two-dimensional space~\cite{PCA}. Specifically, we display a variety of user profiles plotted on a plane across three distinct datasets in Figure~\ref{fig:samples}.
For a comprehensive comparison, we randomly select user profiles categorized as Normal, Random, GSPAttack, and ToDA, where "Normal" refers to the genuine user profiles.
Across the three datasets, it is observed that the user profiles generated via different shilling attacks largely align with the distribution of the Normal category, thus substantiating the imperceptibility.
Upon closer examination, certain outliers, specifically those generated randomly, exhibit a notable deviation from the normal distribution, particularly within the ML-100K and Gowalla datasets. Conversely, the user profiles generated through GSPAttack and ToDA maintain a consistent yet diverse distribution.

%% file: sections/2-Related.tex.bak
\section{Related Work}
We group our work from two perspectives to organize related work, including shilling attacks on recommendation system and diffusion models applied for various domains.

\subsection{Shilling Attacks on Recommendation System}
While significant strides have been made in the development of recommendation systems (RS)~\cite{BPR,NCF, LightGCN, SLMRec}, with a considerable focus on defense mechanisms~\cite{FNCF, AMR, EliMRec}, it is equally crucial to address the vulnerabilities inherent to these systems. One such prevalent vulnerability in RS is the shilling attack~\cite{TrustworthyRS}.
Previous efforts primarily focus on devising optimization functions~\cite{DPAF, PAGR, RevAttack, DPACR, DLAttack} or employing heuristic strategies~\cite{shillingAttack, ReverseAttack, fang2020influence}. However, during the nascent stages of research, attention was drawn towards generative models, such as autoencoder-based~\cite{PCAttack,RAPU-R}, Generative Adversarial Network (GAN)-based~\cite{TriAttack, GSA-GAN, AUSH,HRGAN, LegUP, GSPAttack}, and Reinforcement Learning (RL)-based~\cite{ARG, TDP-CP, KGAttack, CopyAttack, LOKI, Poisonrec} approaches, which operate in an autoregressive manner. Notably, GAN-based methods like LegUP~\cite{LegUP} and GSPAttack~\cite{GSPAttack} demonstrate exceptional attack capability while ensuring imperceptibility. Nevertheless, to attain malicious goals, an additional attack module is frequently integrated, either by utilizing influence functions~\cite{TriAttack} or by formulating additional optimization objectives~\cite{LegUP}. Despite these advancements, more compatible generative models, such as Diffusion models, remain unexplored, lacking a consistent way to achieve both imperceptibility and harmfulness.

\subsection{Diffusion Models}
DMs have demonstrated significant potential, especially in the generation of high-quality synthetic data~\cite{DMs_Survey}, emerging as a sturdy alternative to other generative frameworks such as GANs across a variety of applications~\cite{DM_ImageSynthesis, DM_TopologyOptim, DM_ImageClass}.        
Over the years, a range of diffusion-based generative models have been introduced, embodying core similarities yet distinct implementations. A notable example is Denoising Diffusion Probabilistic Models (DDPMs)~\cite{DDPM}.
Beyond vision-relevant tasks~\cite{DMs_vision_survey}, DMs find extensive utility across diverse domains such as graph~\cite{DMs_graphSurvey} and natural language processing~\cite{Diffuser, Diffuseq}.
Of late, DMs are being leveraged for more complex tasks, exhibiting superior performance in unexplored domains.
In the realm of RS, DMs are integrated to facilitate time-sensitive modeling~\cite{DiffRec} and adeptly user diverse intentions~\cite{DiffuRec}.
Furthermore, within the adversarial attack domain, DMs contribute to the mitigation of sample perturbations~\cite{DM_purification}, generating harmful visual samples~\cite{DiffAttack} and enhancing model robustness~\cite{DM-Improves-AT, DensePure}.
In efforts to guide generated results, substantial strides have been made in text-image generation~\cite{LDM, ControlNet, Diffusionclip}, rendering the process more controllable.
Distinct from preceding endeavors, we showcase the implementation of DMs in thwarting shilling attacks, particularly taking into consideration domain conflicts and target orientation.

%% file: sections/6-Conclusion.tex
\section{Conclusion}

In this paper, we presented a novel shilling attack model, ToDA, by exploiting the potential of DMs. To mitigate the conflicts arising across different domains, we identified the principal challenges to be the benign and narrow focus of DMs, and proposed a Target-oriented Approximator that is seamlessly integrated within the Latent Diffusion Attacker framework. 
And simultaneously we follow the light design principle to mitigate the heavy computational cost in conventional DMs.  
Specifically, we transform user profiles into a latent space to streamline the generation process. Subsequently, we leverage the foundational structure of DMs by injecting noise into these latent features, employing ToA to not only reconstruct the features but also to orient them to be malicious effectively using a cross-attention mechanism in a global view.
Our empirical evaluations underscored the efficacy of ToDA and the rationale of its design. 
Moving forward, we will steer the future focus from four distinct directions: 
1) improving the performance on small datasets;
2) expanding our scope to other recommendation domains, including sequential, multimedia recommendation and \etc;
3) investigating alternative strategies for effectively shilling attacks, such as Large Language Models (LLMs);
and ultimately, 4) designing defensive strategies against such attacks.


%% file: sections/Appendix.tex
\newpage

\section*{Supplemental Materials}

\subsection{Ethics Statement}
In this research, we explore the theoretical and practical implications of using diffusion models for shilling attacks, with our ultimate goal being to develop possible defensive strategies based on the careful analysis of adversarial behaviors. It is crucial to state that this work strictly adheres to ethical guidelines and is conducted with the intent of advancing knowledge and developing preventive measures against such attacks. The step-by-step generation of shilling attacks by ToDA presents a valuable opportunity to understand their mechanisms, which is crucial in developing countermeasures against these attacks. Additionally, defenders can utilize a mix of models (\eg user behavior analysis and anomaly detection) and sampling strategies, such as bagging, to mitigate the impact of these shilling attacks.
We acknowledge the potential for misuse of these techniques and strongly advocate against their application for unethical or illegal purposes. This study does not endorse or facilitate malicious activities; rather, it aims to contribute to the broader understanding of cybersecurity threats and defense mechanisms. Furthermore, all experiments were conducted in controlled environments without real-world impact. We emphasize the importance of ethical conduct in research and the responsibility of the scientific community to use findings for the betterment of society, ensuring that technology advancements do not compromise ethical standards or public trust.

\subsection{Pesudo-code of Training and Inference}

\begin{algorithm}[H]
\caption{Training of ToDA.}
\label{alg:train}
\begin{algorithmic}[1]
\Require user-item interactions $\mathbf{Y}$, target items $\mathcal{T}$ and the number of generated fake users $k$.
\State \textbf{repeat}
\State \quad sample a minibatch of users $\mathcal{U}'$.
\State \quad perform $\mathcal{L}_r$ according to Eq.~\ref{eq:reconstruct}.
\State \textbf{until} converged

\State \textbf{repeat}
\State \quad sample a minibatch of users $\mathcal{U}'$.
\State \quad \textbf{for} all $\mathbf{Y}_u \in \mathbf{Y}_{\mathcal{U}'}$ \textbf{do}: 
\State \quad \quad compute $\mathbf{e}_u = \mathcal{E}(\mathbf{Y}_{u})$.
\State \quad \quad sample $s\sim [1,S]$, $\boldsymbol{\epsilon}_s \sim \mathcal{N}(0,1)$.
\State \quad \quad perform $\mathcal{L}_d$ according to Eq.~\ref{eq:diff_loss}.
\State \textbf{until} converged
\Ensure optimized $\theta$
\end{algorithmic}
\end{algorithm}

\begin{algorithm}[H]
\caption{Inference of ToDA.}
\label{alg:inference}
\begin{algorithmic}[1]
\Require $\theta$ and the interaction history $\mathbf{Y}_u$ of user $u$.
\State compute $\mathbf{e}_{u} = \mathcal{E}(\mathbf{Y}_{u})$.
\State compute $\mathbf{e}_{u;S}=\sqrt{\bar{\alpha}_S}\mathbf{e}_{u} + \sqrt{1 - \bar{\alpha}_S}\boldsymbol{\epsilon}_S$.
\State \textbf{for} $s=S,\dots,1$ \textbf{do}:
\State \quad sample $\boldsymbol{\epsilon} \sim \mathcal{N}(0,1)$ if $s > 1$ else $\boldsymbol{\epsilon}=0$
\State \quad $\mathbf{\hat{e}}_{u;s-1} = \frac{1}{\sqrt{\alpha_s}} (\mathbf{\hat{e}}_{u;s} - \frac{1-\alpha_s}{\sqrt{1-\bar{\alpha}_s}}\boldsymbol{\epsilon}_\theta(\mathbf{\hat{e}}_{u;s}, s)) + \sqrt{\beta_s}\boldsymbol{\epsilon}$.
\State compute $\mathbf{Y}_u^a = \mathcal{D}(\mathbf{\hat{e}}_{u;0})$.
\Ensure crafted user profile $\mathbf{Y}^a_{u}$
\end{algorithmic}
\end{algorithm}

\subsection{Hyper-parameters Analysis}
Following the conventional diffusion models, we tune ToDA with different hyper-parameters to obtain the optimal model for different datasets. Here, we showcase the performance comparison on essential hyper-parameters of diffusion step $S$, the noise range $\beta$, learning rate and L2 regularization.

\begin{figure}
    \centering
    \includegraphics[width=\linewidth]{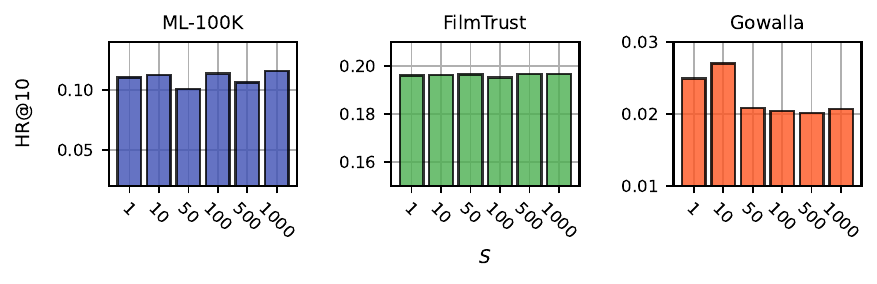}
    
    \caption{The performance comparison in terms of different number of diffusion steps $S$.}
    \label{fig:steps}
    
\end{figure}

\hspace{-5mm}
\textbf{Effect of diffusion step $S$.}
We examine the influence of hyper-parameter $S$ as depicted in Figure~\ref{fig:steps}. Generally, larger steps correspond to enhanced performance on the ML-100K and FilmTrust datasets, albeit with a few exceptions. Contrarily, the trend observed in the Gowalla dataset is distinct, reaching a peak at smaller step values (\ie $S=10$), prior to a marked decline upon incrementing the step size. This empirical evidence suggests that in the context of smaller datasets, a lengthier diffusion procedure may be necessitated, albeit with potential instability. Conversely, for larger datasets like Gowalla, a smaller value of $S$ proves to be adequate.

\begin{figure}
    \centering
    \includegraphics[width=\linewidth]{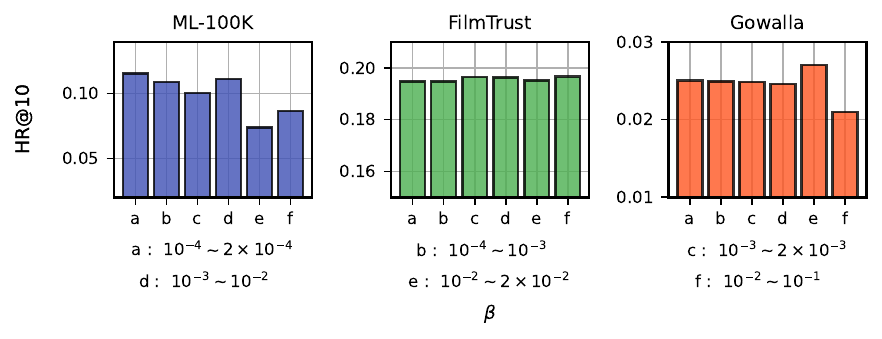}
    
    \caption{The performance comparison in terms of different noise scales.}
    \label{fig:noise}
    
\end{figure}

\hspace{-5mm}
\textbf{Effect of noise scale $\beta$.} 
Throughout each step's diffusion, we add noise to the latent features with different noise scales to investigate its impact to ToDA.
As illustrated in Figure~\ref{fig:noise}, ToDA exhibits different trends to noise scales across different datasets. 
Notably, the choice of noise exerts less influence on Gowalla than the other two datasets. 
Conversely, larger noise magnitudes reveal distinct trends on ML-100K and FilmTrust, with an observed decreasing and increasing trend, respectively.

\begin{figure}
    \centering
    \includegraphics[width=\linewidth]{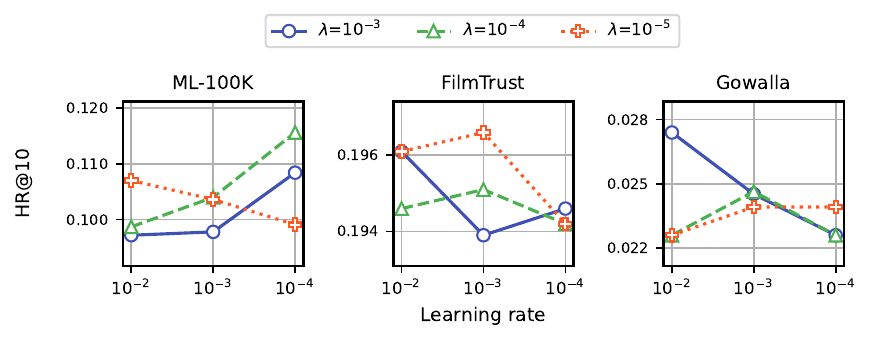}
    
    \caption{The performance comparison in terms of different learning rates and L2 regularization.}
    \label{fig:lr_l2}
    
\end{figure}

\hspace{-5mm}
\textbf{Effect of learning rate and L2 regularization.}
The learning rate, and L2 regularization, collectively influence attack efficiency during the training phase. 
Therefore, we conduct a series of comparative experiments on learning rate and weight decay for three datasets (ML-100K, FilmTrust, and Gowalla) to analyze their impact. The empirical evaluation results are visually presented in Figure~\ref{fig:lr_l2}.
Here $\lambda$ is represented as the strength of L2 regularization (\ie weight decay). 
A discernible trend emerges from the figures corresponding to ML-100K and FilmTrust datasets, particularly at $\lambda=10^{-3}$ and $\lambda=10^{-5}$. Contrarily, the trend manifested in the Gowalla dataset exhibits a marginal difference, showcasing an upswing at $\lambda=10^{-3}$ and a downturn at $\lambda=10^{-5}$.
A notable observation is the different sensitivities exhibited by the three datasets at a specific setting of $\lambda=10^{-4}$.

